\documentclass[thmsa,10pt,a4paper]{article}
\usepackage{sw20lart}

\input tcilatex
\QQQ{Language}{
American English
}

\begin{document}

\author{Maria C. Neacsu}
\title{The magnetic field of the relativistic stars in the 5D approach }
\date{Department of Physics, Technical University ''Gh. Asachi'' Iasi 6600,
Romania }
\maketitle

\begin{abstract}
It is well-known that the 5D equations without sources may be reduced to the
4D ones with sources, provided an appropriate definition for the
energy-momentum tensor of matter in terms of the extra part of the
geometry.The advantage consists on the naturally appearance of gravitational
and electromagnetic fields from this decomposition. With this ansatz an
algorithm is presented, which permits to express the physical parameters in
terms of gauge potentials and scalar field. An explicit form for the
exterior magnetic field of neutron star in terms of the scalar field and the
gauge potentials is deduced for a static, spherically-symmetric metric.
\end{abstract}

\section{Introduction}

Since a great amount of astrophysical objects in Cosmos, like pulsars,
quasars or black holes, are gravitational bodies endowed with magnetic
field, it is of interest to study the nature and behavior of magnetic field
in interaction with gravitation. Classically, the Einstein-Maxwell theory
should be sufficient to describe such cases and, indeed, there exist a set
of exact solutions of those equations representing exterior fields of
gravitational objects endowed with magnetic field \cite{1}. Some of them are
reasonably small, but they do not have the right behavior of the
gravitational field far away from the sources; the other ones are acceptable
in their behavior at infinity , but are rather cumbersome to be studied in
analytic form. Furthermore, this theory actually is not an unification
theory, but rather a superposition one: Einstein plus Maxwell, where the
electromagnetic field appears like a model, there is no explanation for its
existence. Recently \cite{2} it has been shown that vacuum solutions in
scalar-tensor theories are equivalent to solutions of general relativity
with imperfect fluid as source, but the scalar fields do not arise from a
natural framework of unification, instead they are put by hand, as in
inflationary models and are therefore artificial fields.

On the other hand, it is well known that the 5D solutions of Einstein
equations in vacuum when projected into four dimensional space-time
correspond to solutions of Einstein's equations with an energy-momentum
tensor of a gravitational field coupled with an electromagnetic field and
also with a massless scalar field \cite{3}. For theories like Kaluza-Klein
(KK) and Low Energy Superstring (LESS), the electromagnetic field is a
component of a more general field, the existence of gravitation and
electromagnetism following from its decomposition. These theories give rise
to the hypotesis that the magnetic field of some astrophysical objects could
be of fundamental origin and the magnetic field could be a consequence of a
more general scalar-gravito-electromagnetic field. The problem that would
arise is the existence of a scalar field whose interaction is so weak that
it has not been measured till now in gravitational fields like the sun, even
though it would posses a scalar field inherent. Nevertheless, that
attractive or repulsive scalar force could have effects in stronger
gravitational fields like pulsars and this gives rise to the hypothesis that
the magnetic field of some astrophysical objects could be fundamental.

Wesson and Leon \cite{4} demonstrated by algebraic means that the 5D KK
equations without sources may be reduced to the Einstein equations with
sources and proved that the extra part of the 5D geometry could be used
appropriately to define an effective 4D energy-momentum tensor. Hongya and
Wesson obtained that 5D black hole solutions of KK theory can be reduced to
new exact solutions of 4D general relativity in which matter is a
spherically symmetric anisotropic fluid of radiation. Patel and Naresh \cite
{5} obtained from the five-dimensional cylindrically symmetric space-time,
by dimensional reduction, the radiation Friedman- Robertson -Walker flat
model. Matos \cite{6} presented a method for generating exact solutions of
Einstein field equations as harmonic maps using the Chiral formalism in 5D.
These solutions represent exterior fields of a gravitational body with
arbitrary electromagnetic fields and whose gravitational potential posses a
Schwarzschild - like behaviour. He further elaborated a model \cite{7}, \cite
{8}, \cite{9} for the magnetic dipole of static bodies based on a 5D gauge
theory that in 4D corresponds to a massive magnetic dipole coupled to a
massless scalar field. louis O Pimentel \cite{10} and then Montesino and
Matos \cite{11} have obtained an effective energy-momentum tensor using the
4-dimensional part of the 5D Einstein equations with cosmological constant.

In this paper we start from the 5D metric on the principal fibre bundle,
with scalar field and non-vanishing gauge potentials and study the
correspondence between 5D Einstein field equations with cosmological
constant and the 4D Einstein equations with sources. The source of the
gravitation is considered to be a perfect magneto-fluid and from the 5D-4D
reduction yields a natural expression for the magnetic field in terms of the
gauge potentials and of scalar field.

\section{The geometry}

High-dimensional relativity seems to be an elegant way for unifying all
interactions in physics and the first idea has been more and more
transformed from the original suggestion of Kaluza and Klein. In this work
the attention is focused upon the five-dimensional relativity, which unifies
gravitation with electromagnetism.

It is not clear how to interpret physically any higher dimension.
Interpretation of the fifth dimension has been done as a massless scalar
field which can be or not associated with a fluid density \cite{12}, or, in
other works, has been interpreted as a magnetic mass \cite{13}. Other
interpretation has been done as a fifth geometric property which shows up
near horizon. In any case, it is worth to explore ways of interpreting the
properties of the 5D solutions in a 4D word.

The version adopted here considers that the 5D Riemannian space $P$ is a
principal fiber bundle with typical fiber $S^1$, the circle. This version is
more convenient for three reasons: it is a natural generalization of $%
U\left( 1\right) $ - gauge theory (Maxwell theory) to curved space-times,
there is no necessity to recur to the so-called Kaluza-Klein or to the $n$ -
mode ansatz and it is no need to impose any restrictions to the functional
dependence of the metric terms on $P$. The geometry used is shortly explain
bellow.

The formalism of gauge theory \cite{14} is described in the framework of
fibre bundle, which is a collection of elements $\left( P,\pi ,M,G\right) $,
explicitly given by three differentiable manifolds: the principal bundle $P$%
, the base manifold $M$ (usually taken to be the space-time manifold with
metric $g$) and the typical fibre $F$ which embodies the gauge freedom being
a structure Lie group with transition functions that acts on $F$ from the
left and the projection $\pi :P\rightarrow M$ whose inverse image $\pi
^{-1}\left( p\right) \equiv F_p$ is the fibre at $p$. A certain gauge
correspond to a certain section of $P$ and the gauge transformations are
vertical automorphism $f_p:F_p\rightarrow F_p$ that change the section
according to $s\rightarrow f_p\circ s$. The gauge field potentials are given
by the connection forms $\omega $ on $P$ that specifies the way in which a
point of $P$ is to be parallel transported along a curve lying in the base
manifold $M$ and yields a corresponding curvatures or field strengths $%
\Omega $. In practice one uses their section-dependent counterparts $%
s^{*}\omega $ and $\ s^{*}\Omega $ defined on the base manifold. The matter
fields are forms on the base manifold with values in a vector space $U$ of
an associated vector bundle and the elements of the corresponding frame
bundle constitute the reference frames used to decompose the matter fields
with respect to $U$.

Let $U\subset M$ be an open subset of $M$. If $\phi :F\rightarrow U\times F$
is a trivialisation of an open subset of $F$ then the physical quantities
can be mapped into the set $U\times F$ through the trivialisation $\phi $,
which means that the total space $P$ looks locally like the direct product
of $M$ and $F$. Let $\tilde{g}=I^2\hat{\omega}\otimes \hat{\omega}$ be a
metric on $F$ and $\hat{g}=\eta _{AB}\hat{\omega}^A\otimes \hat{\omega}%
^B=\eta _{AB}\hat{\omega}^A\otimes \hat{\omega}^B+I^2\hat{\omega}\otimes 
\hat{\omega}$; $A,B=1,...5$; $a,b=1,...4$ be the metric on $P=H\otimes V$
where $V$ is the vertical space and $H$ its complement. Let $\left\{
e_a\right\} $ be the projection of the complement base $\left\{ \hat{e}%
_a\right\} $ such that $d\pi \left( e_a\right) =e_a$ and $\pi _1$ be the
first projection, $\pi _1:U\times F\rightarrow U$, $\pi _1\left( x,y\right)
=x.$ Thus, because of the identity $\pi =\pi _1\circ \phi $, one finds that
the corresponding base in the trivialisation set $U\times F$ is $d\phi
\left( \left\{ \hat{e}_a,e\right\} \right) =\left\{ e_a-A_a\left( \partial
/\partial y\right) ,\left( \partial /\partial y\right) \right\} $ whose dual
base is $\left\{ \omega ^a,dy+A_a\omega ^a\right\} $. Thus, the
corresponding metric on $U\times F$ will be:

\begin{equation}
\bar{g}=\eta _{ab}\hat{\omega}^a\otimes \hat{\omega}^b+I^2\left( A_a\omega
^a+dy\right) \otimes \left( A_b\omega ^b+dy\right)   \label{1}
\end{equation}
where on observe that $s^{*}\left( \hat{\omega}\right) =A_a\omega ^a.$This
means that $A_a$ are the pullback components of the one-form connection $A$
through a cross section $s$. Since the group $U\left( 1\right) \cong F$ is
acting on $P$, there exist an isometry $Is:P\rightarrow P$, $\left(
x^a,y\right) =\left( x^a,y+2\pi \right) $ such that $Is^{*}\bar{g}=\bar{g}$.
This implies the existence of a Killing-vector $\xi $ and therefore we can
be choose a coordinate system where the metric components of $\bar{g}$ do
not depend on $x^5=y.$ With the gauge-theory philosophy, the action of $%
U\left( 1\right) $ on $P$ means that there are electromagnetic interactions
on $P$, which implies that there is a coordinate system where the metric
components do not depend on  $x^5$.

\section{Space-time field equations}

In the present work the results of \cite{13} are extended with the
assumption of nonvanishing electromagnetic potential. The intention is to
demonstrate that the Einstein equations on $M$ without cosmological constant
and with perfect charged fluid as source, are obtained from the field
equations for vacuum with cosmological constant on the principal fibre
bundle $P\left( 1/JM,U\left( 1\right) \right) $, $J$ being the scalar field
which correspond to the radius of the internal space $U\left( 1\right) $,
whose units and magnitude depends on the particular cases of the
cosmological or astrophysical model \cite{6}. Louis O. Pimentel \cite{10}
used the following identification between the velocity 4-vector and the
scalar field:

\begin{equation}
u_a=\frac{J_{;a}}{\sqrt{-J_{;c}J^{;c}}}  \label{2}
\end{equation}
in order to exprime the energy-impuls tensor in terms of $J$ in the
framework of the $5D$ gravity. If the velocity 4-vector $u_a$ corresponds to
the comoving observers with the fundamental fluid-flow lines, then $u_a$ is
a timelike normalized vector field. In this case, a dependence on the $t$
coordinate of the scalar field will be imposed: $J=J\left( t\right) $, which
means that the scalar field includes the time evolution of the Universe.

The base space $M$ is the space-time of general relativity and we can choose
its metric from the verified metric solutions of the Einstein equations. The
exterior of a spherical, rotating, charged astrophysical object can be
modeled with a static, spherically symmetric metric. Therefore, the metric
will admit two Killing vectors associated with the $(t,\varphi )$
coordinates and the physical quantities will depend only on $(r,\theta )$.
To simplify the calculation, without losing generality, the normalized
gravitational potentials $g_{ab}=(g_1,r^2,r^2\sin ^2\theta ,-g_4)$, the
gauge potentials $A_a=(A_1,A_2,A_3,A_4)$ and the physical parameters will be
chosen to depend only on $r$.

After all those considerations the 5D metric is rather complicated:

\begin{equation}
\bar{g}_{AB}=\left( 
\begin{array}{ccccc}
\frac 1Jg_1+J^2A_1^2 & 0 & 0 & 0 & J^2A_1 \\ 
0 & \frac 1Jr^2+J^2A_2^2 & 0 & 0 & J^2A_2 \\ 
0 & 0 & \frac 1Jr^2\sin ^2\theta +I^2A_3^2 & 0 & J^2A_3 \\ 
0 & 0 & 0 & -\frac 1Jg_4+J^2A_4^2 & J^2A_4 \\ 
J^2A_1 & J^2A_2 & J^2A_3 & J^2A_4 & J^2
\end{array}
\right)  \label{3}
\end{equation}

\begin{eqnarray*}
&& \\
&&
\end{eqnarray*}
and the Einstein field equations on the 5D manifold in vacuum with
cosmological constant $\Lambda $, given by: $\bar{G}_{AB}=\Lambda \bar{g}%
_{AB}$ are cumbersome to calculate (all the calculations were performed
within the GRTensor package \cite{16})

Simplifications could be made to the chose of the Yang-Mills potentials. The
simplest case is for vanishing electromagnetic potential: $A_a=0$. The 4D
form of the Einstein tensor is calculated for the static spherically
symmetric metric:

\begin{equation}
ds^2=g_1dr^2+r^2\left( d\theta ^2+\sin ^2\theta d\varphi ^2\right) -g_4dt^2)
\label{4}
\end{equation}
which for $g_4=1/g_1=g$ is a Schwarzschild-type and for $g_1=a^2/\left(
1-kr^2\right) $, $g_4=1$, is a FRW-type metric.

The extra terms obtained from the identification of the 4D form with the
corresponding part of the 5D Einstein field equation are interpreted to be
the source of the gravity: $G_{ab}=8\pi T_{ab}$. Thus, a nonvanishing
energy-momentum tensor associated with the scalar field $J$ can be defined
(comma means derivation with $t$ ):

\begin{equation}
T_r^r=T_\theta ^\theta =T_\varphi ^\varphi =\frac 1{8\pi }\left( \frac
34\left( \frac{J^{\prime }}J\right) ^2\frac 1{g_1}+\frac \Lambda J\right)
\label{5}
\end{equation}

\begin{equation}
T_t^t=-\frac 1{8\pi }\left( \frac 34\left( \frac{J^{\prime }}J\right)
^2\frac 1{g_1}-\frac \Lambda J\right)  \label{6}
\end{equation}

In this case, it is possible the identification with the corresponding
physical energy-momentum tensor of the perfect fluid:

\begin{equation}
T_{ab}=\left( p+\rho \right) u_au_b+pg_{ab}  \label{7}
\end{equation}
where $u_a$ is the unity velocity vector tangent to the flow lines: $%
u_au_a=-1$.

From this comparison, the pressure and the density of energy can be
expressed in term of the scalar field:

\begin{equation}
\rho =\frac 34\left( \frac{J^{\prime }}J\right) ^2\frac 1{g_4}-\frac \Lambda
J  \label{8}
\end{equation}
\begin{equation}
p=\frac 34\left( \frac{J^{\prime }}J\right) ^2\frac 1{g_4}+\frac \Lambda J
\label{9}
\end{equation}
where the gravitational constant is $\Lambda =G_{yy}$.

The case of nonvanishing Yang-Mills potentials will be considered further.
In the rest frame of the spherical charge distribution, if the gauge has the
form: $A_a=\left( 0,0,A_3,0\right) $, the electromagnetic tensor: $%
F_{ab}=A_{b,a}-A_{a,b}$ will admit only one nonvanishing component
corresponding to the magnetic field: $B_a=1/2\varepsilon
^{abcd}u_bF_{cd}=(0,B_\theta ,0,0)$.

The introduction of the magnetic potential in the 5D form of the metric has
been made with the intention of finding an identification between the 4D
part of Einstein tensor obtained from the reduction 5D-4D and the
energy-momentum tensor which describes the external behavior of a magnetized
astrophysical object. The existence of the azimuthal magnetic field is
realistic for the neutron star, this being the only stable configuration for
a stationary axisymmetric flow of the internal stellar plasma in the MHD
approximation.

The energy-momentum tensor is constituted with the perfect fluid tensor and
the magnetic part of the electro-magnetic tensor, yielding the complete form
of the perfect magneto-fluid tensor:

\begin{equation}
T_{ab}=\left( p+\rho +\frac{B^2}{4\pi }\right) u_au_b+\left( p+\frac{B^2}{%
8\pi }\right) g_{ab}-\frac{B_aB_b}{4\pi }  \label{10}
\end{equation}
which for the chosen metric has the components:

\begin{equation}
T_b^a=\left( 
\begin{array}{cccc}
p+\frac{B_\theta ^2}{8\pi r^2} & 0 & 0 & 0 \\ 
0 & p-\frac{B_\theta ^2}{8\pi r^2} & 0 & 0 \\ 
0 & 0 & p+\frac{B_\theta ^2}{8\pi r^2} & 0 \\ 
0 & 0 & 0 & -\rho -\frac{B_\theta ^2}{8\pi r^2}
\end{array}
\right)  \label{11}
\end{equation}
\[
\]

In terms of the components introduced by the fifth dimension, the tensor $%
(11)$ has the form (dot means derivation with $r$):

\begin{equation}
T_r^r=\frac 1{8\pi }\left( \frac 34\left( \frac{J^{\prime }}J\right) ^2\frac
1{g_4}+\frac \Lambda J+\frac{J^3\left( \dot{A}\right) ^2}{4r^2g_1\sin
^2\theta }\right)  \label{12}
\end{equation}

\begin{equation}
T_\varphi ^\varphi =\frac 1{8\pi }\left( \frac 34\left( \frac{J^{\prime }}%
J\right) ^2\frac 1{g_4}+\frac \Lambda J-\frac{J^3\left( \dot{A}\right) ^2}{%
4r^2g_1\sin ^2\theta }\right)  \label{13}
\end{equation}
\begin{equation}
T_\varphi ^\varphi =\frac 1{8\pi }\left( \frac 34\left( \frac{J^{\prime }}%
J\right) ^2\frac 1{g_4}+\frac \Lambda J+\frac{J^3\left( \dot{A}\right) ^2}{%
4r^2\sin ^2\theta }\right)  \label{14}
\end{equation}
\begin{equation}
T_t^t=\frac 1{8\pi }\left( -\frac 34\left( \frac{J^{\prime }}J\right)
^2\frac 1{g_4}+\frac \Lambda J-\frac{J^3\left( \dot{A}\right) ^2}{%
4r^2g_1\sin ^2\theta }\right)  \label{15}
\end{equation}
From the comparison of $(11)$, with $(12-14)$ the expression for the
magnetic field is naturally obtained:

\begin{equation}
B_\theta =\frac{J^3\left( \dot{A}\right) ^2}{4g_1\sin ^2\theta }  \label{16}
\end{equation}
The equation $(16)$ clearly describes the universal constitution of the
magnetic field and its essential contribution to metric geometry of the
space-time.

The expression for the cosmological constant, derived from the fifth
component of the Einstein tensor, is rather complicated, but the nondiagonal
component $G_{\varphi y}$ provides with an interesting relation:

\begin{equation}
\frac{\dot{A}}A\left( \frac{\dot{g}_4}{g_4}-\frac{\dot{g}_1}{g_1}\right) =2%
\frac{\ddot{A}}A  \label{17}
\end{equation}
which can be easily integrated and a formal expression for the metric
potential in terms of the scalar and gauge field is deduced:

\begin{equation}
\frac{g_1}{g_4}=\dot{A}^2  \label{18}
\end{equation}

The expession $(18)$ is then substituten in $(16)$ and yields an interesting
formulae that gives the time-dependence of the magnetic field:

\begin{equation}
B_\theta =\frac{J^3}{4g_4\sin ^2\theta }  \label{19}
\end{equation}

For $g_4=1/g_1$ (Schwarzschild case) we obtain: $g_1=\dot{A}$, and for $%
g_4=1 $ (FRW case), we have: $g_1=\dot{A}^2$. Considering those metric
functions, the magnetic field becomes in the first case:

\begin{equation}
B_\theta =\frac{J^3\dot{A}}{4\sin ^2\theta }  \label{20}
\end{equation}
and in the second:

\begin{equation}
B_\theta =\frac{J^3}{4\sin ^2\theta }  \label{21}
\end{equation}

\section{Conclusions}

In this paper is has been shown that the field equations with cosmological
constant $\Lambda $ on the principal fibre bundle $P\left( 1/JM,U\left(
1\right) \right) $ with nonvanishing magnetic potential, can lead to the
energy-momentum tensor of a perfect magneto-fluid endowed with azimuthal
magnetic field, which is the case of a neutron star [17]. The magnetic field
yields from a natural decomposition in direct connection with the fifth
dimension and this is of evidence for the fundamental role of the magnetic
field to the constitution of matter and of the Universe.

The study of the above problems involve a lot of interesting aspects besides
the ones analyzed here. It can be pointed out, for instance, the study of
the 4D energy momentum tensor which corresponds to this solution; the study
of the behavior of test particles and of future interest is the
generalization of this solution to the stationary case in order to modelate
in a more realistic way a rotating astrophysical object. It is also
convenient to make a deeper investigation on the relationship between this
model and the string theory for low energies.

\end{document}